\documentclass[12pt]{iopart}
\usepackage{graphicx,epsfig}
\usepackage{iopams}

\expandafter\let\csname equation*\endcsname\relax
\expandafter\let\csname endequation*\endcsname\relax

\usepackage{soul}

\usepackage{amsmath}
\usepackage{amssymb}
\usepackage{color}
\usepackage{xcolor}
\usepackage[catalan]{babel}

\usepackage{slashbox}

\usepackage{fancyhdr}

\usepackage[T1]{fontenc}
\usepackage[utf8]{inputenc}

\newcommand{\beq}{\begin{equation}} 
\newcommand{\eeq}{\end{equation}} 

\newcommand{\ket}[1]{\left|#1\right\rangle} %|"cosa">
 \newcommand{\bra}[1]{\left\langle#1\right|} %<"cosa"|
 \newcommand{\braket}[2]{\left< #1 \vphantom{#2} \right|
  \left. #2 \vphantom{#1} \right>} % for Dirac brackets
 \newcommand{\expval}[1]{\left\langle#1\right\rangle} %<"cosa">

\begin{document}
%\bibliographystyle{plain}

%%%%%%%%%%%%%%%%%%%%%%%%%%%%%%%%%%%%%%

%%%%%%%%%%%%%%%%%%%%%%%%%%%%%%%%%%%%%%
%\pagestyle{fancy}
%\lhead{\bf Exact diagonalization of Bose-Hubbard Models in small lattices}
%\rhead{David Raventós i Ribera}
%\lhead{}
%\rhead{}
%\lfoot{Treball de Fi de M\`aster.}
%\rfoot{Barcelona, June 2014}
%%%%%%%%%%%%%%%%%%%%%%%%%%%%%%%%%%%%%%

\title{Cold bosons in optical lattices: a tutorial for Exact Diagonalization}

\author{David Raventós $^1$ \footnote{david.raventos@icfo.eu},  Tobias Gra{\ss}$^{1,2}$, Maciej Lewenstein$^{1,3}$, and Bruno Juli{\'a}-D{\'i}az$^{1,4,5}$}

\address{$^1$ ICFO-Institut de Ciencies Fotoniques, The Barcelona Institute of Science and Technology, 08860 Castelldefels (Barcelona), Spain}
\address{$^2$ Joint Quantum Institute, University of Maryland, College Park, MD 20742, U.S.A.}
\address{$^3$ICREA, Passeig de Llu\'is Companys, 23, 08010 Barcelona, Spain}
\address{$^4$ Departament de Física Qu\`antica i Astrof\'isica, Facultat 
de Física, 
Universitat de Barcelona, Barcelona 08028, Spain}
\address{$^5$ Institut de Ci\`encies del Cosmos, Universitat de 
Barcelona, ICCUB, 
Martí i Franqu\`es 1, Barcelona 08028, Spain}
\begin{abstract}
Exact diagonalization techniques are a powerful method for studying
many-body problems. Here, we apply this method to systems of few bosons in an optical lattice, and use it to
demonstrate the emergence of interesting quantum phenomena like fragmentation and coherence. 
Starting with a standard Bose-Hubbard Hamiltonian, we first revise the characterization of the superfluid 
to Mott insulator transitions. We then consider an inhomogeneous lattice, where one potential minimum is made much deeper 
than the others. The Mott insulator phase due to repulsive on-site interactions then competes with the trapping of all atoms in the deep potential. Finally, we turn our attention to attractively interacting 
systems, and discuss the appearance of strongly correlated phases and the onset 
of localization for a slightly biased lattice. The article is intended to 
serve as a tutorial for exact diagonalization of Bose-Hubbard models.
\end{abstract}

\submitted{\jpb}

\maketitle

%\ioptwocol

%\tableofcontents

%%%%%%%%%%%%%%%%%%%%%%%%%%%%%%%%%%%%%%%%%%%%%%%%%%%%%%%%%%%%%%%%%%%%%%%%%%%
\section{Introduction}

The Bose-Hubbard model (BHM), originally introduced in order to describe 
different phenomena in condensed matter physics~\cite{PhysRevB.40.546}, 
has gained new impact in the field of quantum gases~\cite{lewenstein2012ultracold}, 
following the experimental realization of the model in a setup with cold atoms 
in optical lattices~\cite{citeulike:890627}. In particular, the prediction of 
a phase transition from a superfluid (SF) to a Mott insulator (MI) has been 
confirmed. The origin of this transition is genuinely quantum, that is, it is 
driven by quantum fluctuations, which are controlled by the Hamiltonian parameters, 
interaction and hopping strength, and which are present also at zero temperature. 

The advantages offered by cold atoms for studying quantum phase transitions are
clear. First, in these systems, high isolation from the surrounding environment 
is achievable. There have been recent advances in producing different sort 
of lattice configurations, determining the Hamiltonian parameters. Second, atom-atom 
interactions are tunable via Feshbach resonances. These properties allow one to 
use ultracold atomic systems as quantum simulators of theoretical models that 
are not tractable with classical computers. Although different techniques are 
able to capture ground state properties of the Bose-Hubbard Hamiltonian, the 
solution of the full model, that is complete spectrum and eigenstates, appears 
to be intractable with classical techniques. Exact diagonalization techniques, 
which in principle allow one to solve the full problem with high accuracy, 
suffer from the clear shortcoming of being restricted to fairly small many-body 
quantum systems~\cite{PhysRevLett.98.110601}.

%Introducing the technical details, there are also
%other works with this aim~\cite{0143-0807-31-3-016}.

Several approaches have been used to study the BHM: Bogoliubov 
techniques at small interactions~\cite{PhysRevA.63.053601}, 
perturbative ones at large interactions~\cite{gel90,PhysRevB.53.2691}, 
Gutzwiller mean-field approaches~\cite{PhysRevB.44.10328,PhysRevLett.81.3108},
field-theoretic studies~\cite{PhysRevA.79.013614,andre,grass}, etc. 
Ground state properties can be studied by means of DMRG methods~\cite{RevModPhys.77.259, PhysRevA.87.043606} 
and Quantum Monte-Carlo techniques~\cite{worm}. 

While the phase boundary between the Mott insulating phase and the superfluid 
phase is well-defined in the thermodynamic limit, where symmetry-breaking gives 
rise to a non-zero order parameter, the situation is less unique for finite 
systems. In particular, as reviewed in Ref.~\cite{RevModPhys.83.1405} and also 
pointed out in~\cite{PhysRevA.87.043606}, there is still uncertainty on the 
precise value of the transition from Mott to superfluid in 1D systems. In 
particular quantum Monte-Carlo studies have produced slightly disagreeing results 
on the critical value of the parameters~\cite{PhysRevLett.65.1765,PhysRevB.46.9051,kashur}.
In view of this, further study of the Mott transition is needed, using different 
techniques and applying different definitions. 
Here, exact methods allow to extract quantities not reachable by means of other methods, such as eigenstates, eigenenergies and the Entanglement spectrum.

In this work we consider small lattices which we study using exact diagonalization (ED). 
We apply and compare different signatures of the MI-SF transition:
Given the full ground state of the system, a simple figure of merit is the overlap between the numerical solution 
and analytical trial states for the Mott and the SF phase. To capture the phase boundary more accurately, we extract the single-particle insulating gap from the energy spectra at different numbers of atoms. Performing a finite 
size scaling, we determine the parameters for which the gap would close in the thermodynamic limit, indicating the transition to the superfluid phase.

Interesting new phenomena are brought into the problem by a simple modification 
of the model, assuming a lattice with one highly biased site attracting the
atoms. This gives rise to a series of quantum phase transitions upon changing
the lattice depth: For certain values, number fluctuations in the system become
strong while the average number of particles on the biased site is decreased by
one.

Finally, we consider the case of attractive interactions. Similarly to the
two-site case discussed in Refs.~\cite{PhysRevA.57.1208,PhysRevA.81.023615},
strong fragmentation is found in the ground state of the system for a small
attractive interaction. Direct diagonalization allows us to quantitatively 
discuss the appearance of many-body correlations in the ground state. 
Considering a slightly biased lattice, we study the onset of localization 
in the system as the attraction is increased. 

The present manuscript is also intended to provide a detailed, tutorial like, 
description of the methods employed to perform the exact diagonalization 
of the model. Our work complements other tutorial like ones, like reference~\cite{0143-0807-31-3-016}, 
as we also incorporate a state-of-the-art discussion of the definition 
of the transition between the MI and superfluid phases. 

This work is organized as follows: The BHM is introduced in Sec.~\ref{sec2}.
In Sec.~\ref{sec2.1}, we introduce different quantities used to 
characterize the system behaviour, such as eigenvalues of the one body density 
matrix, and the populations of the Fock states. They allow us to discern if the 
system is condensed and to measure its spatial correlations. We also define 
different entropies in order to capture important properties about the system 
with a single scalar value.  In Sec.~\ref{sec2.2}, we present the phases 
exhibited by the Bose-Hubbard model. In Sec.~\ref{sec3} we explain the exact
diagonalization techniques used together with a detailed description of how to perform them.
In Sec.~\ref{sec4} we present the $U/t$ value at which the MI-SF phase transition takes place for the BHM at filling 1, applying 
several finite size studies to our exact diagonalization results. In Sec.~\ref{sec5}, 
we go beyond the standard BHM: In Sec.~\ref{sec5.1}, we study an inhomogeneous 
lattice, and observe several transitions as the hopping and/or interaction 
strengths are varied, and in Sec.~\ref{sec5.2}, we turn to attractive interactions, 
focusing on the appearance of correlated states. 
In Sec.~\ref{sec6}, the reader is briefly introduced to the treatment
 of quantum Hall effects with Exact Diagonalization.
Conclusions are given in Sec.~\ref{sec7}.

%%%%%%%%%%%%%%%%%%%%%%%%%%%%%%%%

\section{The Bose-Hubbard model and its characterization}
\label{sec2}

We start considering the standard Bose-Hubbard model which contains two terms: 
the hopping term, which allows the exchange of particles between the sites, 
related to the kinetic energy, and the on-site interaction term, which can 
be repulsive or attractive. The Hamiltonian of the model reads,
\begin{equation}
\hat{\mathcal{H}}= 
-\sum_{j \neq k}^{M}  t_{k,j} \hat{a}^{\dagger}_j \hat{a}_k 
+ \frac{U}{2} \sum_{i=1}^{M} \hat{n}_i(\hat{n}_i -1) 
\equiv 
\sum_{j\neq k}^M 
\hat{\mathcal{T}}_{k,j} + 
\sum_{i=1}^M \hat{\mathcal{U}}_i
\label{eq:Ham_phi_zero}
\end{equation}
where $\hat{a}_j^{\dagger}$ ($\hat{a}_j$) creates (annihilates) one particle in 
the $j$th site and $\hat{n}_i=\hat{a}_i^{\dagger}\hat{a}_i$ is the number of 
particles operator in the $i$th site, being $M$ the number of sites. A 
convenient finite basis, with a fixed number of particles $N$, is given by the states 
of the Fock space restricted to $N$ particles,
\begin{equation}
\ket{\beta}\equiv\ket{n_1^{\beta} , n_2^{\beta} , \cdots, n_{M}^{\beta}}
\equiv {\frac{1}{\sqrt{n_1! n_2! \dots n_M!}}} 
\left(\hat{a}_1^\dagger\right)^{n_1} \left(\hat{a}_2^\dagger\right)^{n_2} 
...
\left(\hat{a}_M^\dagger\right)^{n_M} 
|{\rm vac}\rangle  
\label{eq:FocSt}
\end{equation}
where $n_i^{\beta}$ is the number of bosons at the $i$th site in the state 
$\ket{\beta}$, and $\beta$ is the labelling of the Fock states. Since the 
number of bosons $N$ in the system is fixed, $n_i^{\beta}$ satisfies $\sum_i^{M}{n_i^{\beta}}=N$ 
for any state $\ket{\beta}$. Arbitrary states can be written in this orthogonal 
basis, 
\begin{equation}
\ket{\Phi}=\sum_{\beta}^{{\cal N}_N^M}{c_{\beta} \ket{\beta}} \,,
\label{eq:FocSt_ba}
\end{equation}
with $c_{\beta} \in \mathbb{C}$.
For total number of bosons $N$ and sites $M$ there are ${\cal N}_N^M$ Fock states 
in the basis. This number is the number of ways of placing $N$ particles in $M$ 
sites, see Table~\ref{tab:hss}, 
\begin{equation}
{\cal N}_N^M=\binom{N+M-1}{N}
=\frac{\left(N+M-1 \right)!}{N!\left(M-1\right)!}\,.
\label{eq:N_bas}
\end{equation}
If the particles were fermions instead of bosons, the number of basis states 
is, 
\begin{equation}
{\cal N}_N^{M,\rm fermions} = \binom{M}{N} \,.
\end{equation} 

\begin{table}
\centering
\scalebox{0.7}{\begin{tabular}{l|rrrrrrrrrrrrr}
\backslashbox{$N$}{$M$}&1&2   & 3 & 4 & 5 & 6 & 7 & 8 & 9 & 10 & 11 & 12 & 13 \\
\hline
1& 1 & 2 & 3 & 4 & 5 & 6 & 7 & 8 & 9 & 10 & 11 & 12 & 13 \\
2& 1 & 3 & 6 & 10 & 15 & 21 & 28 & 36 & 45 & 55 & 66 & 78 & 91 \\
3& 1 & 4 & 10 & 20 & 35 & 56 & 84 & 120 & 165 & 220 & 286 & 364 & 455 \\
4& 1 & 5 & 15 & 35 & 70 & 126 & 210 & 330 & 495 & 715 & 1001 & 1365 & 1820 \\
5& 1 & 6 & 21 & 56 & 126 & 252 & 462 & 792 & 1287 & 2002 & 3003 & 4368 & 6188 \\
6& 1 & 7 & 28 & 84 & 210 & 462 & 924 & 1716 & 3003 & 5005 & 8008 & 12376 & 18564 \\
7& 1 & 8 & 36 & 120 & 330 & 792 & 1716 & 3432 & 6435 & 11440 & 19448 & 31824 & 50388 \\
8& 1 & 9 & 45 & 165 & 495 & 1287 & 3003 & 6435 & 12870 & 24310 & 43758 & 75582 & 125970 \\
9& 1 & 10 & 55 & 220 & 715 & 2002 & 5005 & 11440 & 24310 & 48620 & 92378 & 167960 & 293930 \\
10& 1 & 11 & 66 & 286 & 1001 & 3003 & 8008 & 19448 & 43758 & 92378 & 184756 & 352716 & 646646 \\
11& 1 & 12 & 78 & 364 & 1365 & 4368 & 12376 & 31824 & 75582 & 167960 & 352716 & 705432 & 1352078 \\
12& 1 & 13 & 91 & 455 & 1820 & 6188 & 18564 & 50388 & 125970 & 293930 & 646646 & 1352078 & 2704156 \\
13& 1 & 14 & 105 & 560 & 2380 & 8568 & 27132 & 77520 & 203490 & 497420 & 1144066 & 2496144 & 5200300 \\
\end{tabular}}
\caption{Size of the Hilbert space for $N$ bosons in $M$ sites, ${\cal N}_N^M$ 
for $N,M=1,\dots,13$.\label{tab:hss}} 
\end{table}

\subsection{Useful quantities}
\label{sec2.1}
Let us introduce some quantities that we will use in this work to discuss the characterization of the BHM.

\subsubsection{Fragmentation in the ultracold gas.}
\label{sec2.1.1}

The generalization of the concept of Bose-Einstein condensation to interacting 
systems was introduced by Penrose and Onsager~\cite{PhysRev.104.576,doi:10.1080/14786445108560954}. 
They established a condensation criterion in terms of the one-body density matrix 
(OBDM),
\begin{equation}
\rho ^{\left( 1 \right)}\left( {\bf r},{\bf r^{\prime}} \right) 
= \expval{\psi^{\dagger}\left( {\bf r^{\prime}} \right)\psi \left( {\bf r } \right)},
\label{eq:OBDM_r}
\end{equation}
where the field operator $\psi^{\dagger}$ creates a boson at position ${\bf r}$ and 
$\expval{\cdots}$ is the thermal average at temperature $T$. Since $\rho ^{\left( 1 \right)}$ 
is a Hermitian matrix, it can be diagonalized. The eigenvectors are termed natural orbitals, 
and the eigenvalues are their corresponding populations.

The way to find out if a given state is condensed involves the computation of the 
OBDM and its diagonalization in order to study the size of the populations 
of its eigenstates. In second quantization, the definition of the OBDM $\rho_{k,l}$ of 
a state $\ket{\Phi}$ is,
\begin{equation}
\rho_{k,l}=\bra{\Phi}\hat{a}^{\dagger}_l \hat{a}_k\ket{\Phi}\,.
\label{eq:OBDM_Def}
\end{equation}
But writing the state $\ket{\Phi}$ as in Eq.~(\ref{eq:FocSt_ba}), we explicitly get,
\begin{equation}
\rho_{k,l}=\sum_{\alpha,\beta}^{{\cal N}_N^M}{c_{\alpha}^*c_{\beta} 
\bra{\alpha}\hat{a}^{\dagger}_l \hat{a}_k \ket{\beta}}\,.
\label{eq:OBDM}
\end{equation}
From the diagonalization of the OBDM in an arbitrary basis, one obtains,
\begin{equation}
\rho_{i,j}=n^{\rm OBDM}_i\delta_{i,j}\,,
\label{eq:OBDM_diag}
\end{equation}
where $n^{\rm OBDM}_i$ is the $i$th largest eigenvalue of the OBDM.

In order to simplify the information given by the eigenvalues of the OBDM of a 
given state, we introduce an entropy based on the von Neumann one, $S_1$, which 
will be used in the following. It is defined as,
\begin{equation}
S_1=-\sum_{i}^{M}{p_i\ln{p_i}}\,,
\label{eq:S1}
\end{equation}
with $p_i=n^{\rm OBDM}_i/N$ the normalized eigenvalues of the OBDM. So, $\sum_{i}{p_i}=1$. 
The minimum of $S_1$ is $0$ and corresponds to $p_i=\delta_{i,1}$. The entropy $S_1$ has 
a maximum which equals $\ln{M}$ when $p_i=1/M,~\forall i$. So, its maximum value corresponds 
to a uniform probability distribution (fragmented condensate~\cite{PhysRevA.74.033612}), 
whereas the minimum corresponds to a Kronecker-$\delta$ distribution, full condensation. 
In all computations, the entropy has been divided by its maximum value, $\ln{M}$, in order 
get a non-extensive quantity, bounded by $0$ and $1$.

The entropy $S_1$ measures condensation, as defined by the Penrose-Onsager criterion. 
When the value is $0$, the system is condensed. When it is $\ln{M}$, it is completely 
fragmented. When the value is the logarithm of a certain integer $r$, the state is 
fragmented in $r$ states.

\subsubsection{Spatial correlations from Fock-space coefficients.}
\label{sec2.1.2}

In order to quantify the correlations between the particles on different sites, we 
take advantage from the fact that our Fock basis builds on spatially localized single 
particle states. We define a second entropy $S_D$, which measures the clustering of 
particles in the Fock space,
\begin{equation}
S_D=-\sum_{\beta}^{{\cal N}_N^M}{{\left| c_{\beta} \right|}^2\ln{{\left| c_{\beta} \right|}^2}}\,,
\label{eq:SB}
\end{equation}
where $c_{\beta}$ are the coefficients of the decomposition of a given state into the 
Fock basis $\ket{\beta}$, Eq.~(\ref{eq:FocSt_ba}). In the same way as the entropy 
$S_1$ allowed us to distinguish between condensed and fragmented states, the entropy 
$S_D$ distinguishes between many-body states which are represented by a single Fock 
state ($S_D=0$), and superpositions of many Fock states ($S_D>0$). Apparently, if only 
few Fock states contribute to a many-body state, there is a high amount of spatial 
correlations in the system, which thus can be captured by the value of $S_D$. The 
entropy $S_D$ is the von Neumann entropy of the diagonal ensemble after tracing 
off one site. This means that it provides the von Neumann entropy after a long-term 
time evolution in a local Hamiltonian $\hat{\cal H} = \sum_i \epsilon_i \hat{n_i}$, 
with $\epsilon_i$ local energies. Note that in the case of solely two-sites, the 
entropy $S_D$ coincides with the left-right bipartite entropy~\cite{PhysRevA.81.023615}.

\subsection{Phases of the BH model}
\label{sec2.2}

The homogeneous case of the Hamiltonian (\ref{eq:Ham_phi_zero}), with $t_{k,j} = t$, becomes exactly solvable in two limiting 
cases: $t/U=0$ and $t/U \rightarrow \infty$. We take ground states in these two 
cases as analytical trial states for the two quantum phases exhibited by the model: 
the non-interacting limit provides a trial state for the SF phase, while the system 
without hopping yields a trial state for the MI phase.

\subsubsection{Mott Insulator regime.}
\label{sec2.2.1}

When $t/U\rightarrow0$ with $U>0$, the system is dominated by the repulsive interactions, 
and it minimizes energy by reducing the number of pairs in each site. So, the GS of 
the system is a state with $q \equiv N/M$ particles on each site, where $q$ is a positive 
integer, i.~e., a Mott insulator state. This corresponds to one many-body state of 
the Fock basis and it reads,
\begin{equation}
\ket{\Phi_{\rm MI}(q)}
=\prod_{i=1}^{M}{\frac{(\hat{a}_i^{\dagger})^q}{\sqrt{q!}}}\ket{0}
=\ket{q \cdots q} \,.
\label{eq:MI}
\end{equation}

The first excited state looks like a MI state where a particle has been annihilated 
in one site and created in a different site, i.~e., it is a quasiparticle-quasihole 
excitation of the MI state. When the particle is created in the $i$th site and the 
hole is localized in the $j$th one, the first excited state reads,
\begin{equation}
\ket{\Phi_{\rm MI}(q)}^{(1)}=
\frac{1}{q}{\hat{a}^{\dagger}_i \hat{a}_j}\ket{\Phi_{\rm MI}(q)}.
\label{eq:FE_MI}
\end{equation}

The Mott insulator is an insulator in the sense that the ``transport'' of one particle 
from one site to another costs a finite amount of energy (the energy gap $\Delta E$). 
In the MI state, when $q$ particles are in one site, the value of the interaction 
term in that site is $(U/2)q(q-1)$. When in the MI state, a particle hops from one site 
to another, the value on the interaction term is $(U/2)(q-1)(q-2)$ in the site 
where the particle comes from and $(U/2)(q+1)q$ in the site where the particle goes. 
This situation coincides with the first excitation of the MI state. So, the energy 
difference of the MI state and its excitation is,
\begin{equation}
\begin{split}
\Delta E &= \frac{U}{2} \left[ (q-1)(q-2)+(q+1)q -2q(q-1) \right] = U.
\end{split}
\label{eq:Gap_MI}
\end{equation}
Thus, the MI phase has a characteristic energy gap $\Delta E=U$ in the energy spectrum 
which separates the ground state from the excitations.

We consider systems at filling one, that is, $q=N/M=1$. In the MI phase, there is 
one particle in each site and $S_1=\log{M}$. Due to the fact that in this phase 
the GS coincides with a single Fock state, $S_D$ is zero. Since the number of 
particles $q$ in each site is a well-defined integer, there are no fluctuations 
on the on-site number of particles in the Mott phase. The MI phase also has a 
finite correlation length $\xi$, defined in $\expval{a_i a_j} - \expval{a_i} \expval{a_j} \propto {\rm e}^{-\left| {\bf r}_i - {\bf r}_j \right|/\xi}$ as a measure of the spatial range of pair correlations.

\subsubsection{Superfluid regime.}
\label{sec2.2.2}

When $U/t \rightarrow 0$, the hopping rules the system and each particle becomes 
completely delocalized over all sites of the lattice. So, we can write the single 
particle state as,
\begin{equation}
\ket{\phi_{\rm sp}}=\frac{1}{\sqrt{M}} 
\sum_{i=1}^{M}\hat{a}_i^{\dagger}\ket{0}\,.
\label{eq:SF_SPWF}
\end{equation}
Since there are no interactions, the state of the whole system is a properly 
symmetrized product of the single particle state up to the number of particles. 
So,
\begin{equation}
\ket{\Phi_{\rm SF}}
=\frac{1}{\sqrt{N!}}\left[\frac{1}{\sqrt{M}} 
\sum_{i=1}^{M}\hat{a}_i^{\dagger}\right]^{N}\ket{0} \,.
\label{eq:SF}
\end{equation}
Then, the squared coefficients of the decomposition of the SF state into the Fock 
basis follow a poissonian distribution in the sense that its variance 
${\rm Var}\left({\left| c_{\beta} \right|}^2\right)$ coincides with its mean 
$\expval{{\left| c_{\beta} \right|}^2}$~\cite{citeulike:890627}.

The SF state is characterized by a vanishing gap (since there is no interaction, 
the only contribution to the gap comes from the hopping term), large fluctuations 
in the on-site number of particles and a divergent correlation function. In the SF 
phase, all particles are delocalized, that is, each one of them has the same 
probability of presence in all sites of the lattice, without interacting with each other. 
Since all the particles in the system have the same single particle wavefunction, 
the system is condensed and so, $S_1=0$. The SF state involves many Fock states 
with a non-uniform distribution. The entropy $S_D$, defined in 
Eq.~(\ref{eq:SB}), is larger than in the Mott phase, but it will never equal $1$ 
because the distribution is not uniform. Increasing the number of particles in the
system, the value of the entropy $S_D$ in the SF phase decreases. In contrast to 
$S_1$, the entropy $S_D$  does not exhibit an extremal value. In Sec.~\ref{sec5}, 
we will encounter cases where the distribution of coefficients is closer to a uniform distribution, 
giving rise to even larger values of $S_D$.

\section{Exact diagonalization}
\label{sec3}

Let us depiece how we have performed the exact diagonalization of the Hamiltonian, 
Eq.~(\ref{eq:Ham_phi_zero}). The same procedure may be applied to many models involving 
particles with bosonic and/or fermionic statistics.

Exact diagonalization is the straightforward way to obtain the eigenvalues 
and eigenvectors of a Hamiltonian. Naively, we first need the Hamiltonian 
written in matrix form in a particular basis of states. The apparent drawback is the 
fast growth of the dimension of this matrix, defined by the size of the basis, 
see Table~\ref{tab:hss}. In general, obtaining the full spectrum of the Hamiltonian, eigenvectors and 
eigenvalues, requires a number of operations which scales as $\left( {\cal N}_N^M \right) ^3$. 
This makes the problem already intractable for fairly small quantum systems, and strictly impossible for larger ones.

Once the Hamiltonian matrix (or its action on arbitrary state vectors) is known, 
there are two classes of algorithms, direct and iterative methods, which can be 
used to completely or partially diagonalize a matrix, that is to find (at least) 
some of its eigenvalues and eigenvectors:
\begin{itemize}
\item {\bf Direct methods} perform 
similarity transformations to the hermitian (non-hermitian) matrix of interest until it is written in 
a reduced form. Hermitian matrices (general non-Hermitian) are reduced to symmetric tridiagonal (upper Hessenberg) matrices. Once the matrix of interest is in the reduced form, it can be eigendecomposed in an efficient way with LU (QR) decomposition for hermitian (non-hermitian) matrices.

\item In the {\bf iterative projection methods}, the matrix operator is applied 
to a set of trial vectors, approximations to the eigenvalues are obtained from 
subspaces of lower dimension, and the iteration is continued until convergence 
is reached. 
Notice that they are able to approximate a number of eigenvalues and eigenvectors
 without any need to solve the entire system. Despite some of them are able to solve the entire system, 
it is not practical in most applications, due to a much larger number of operations than required by direct methods.
\end{itemize}

The direct methods are the only ones that are able to truly diagonalize 
a matrix, up to rounding machine errors, while the second ones obtain 
approximate partial solutions of increasing precision in an iterative way. On the 
other hand, direct methods require enough memory to store the full Hamiltonian 
and the similarity matrix, while iterative methods only need storage for a few state vectors.
 Matrix elements needed to compute the action of the matrix onto a state vector can either
 be determined on the fly, or stored in a less costly sparse-matrix format.

In our case, we have used an iteration projection method for sparse, hermitian 
problems: the Lanczos algorithm. In order to implement it, a number of libraries 
are publicly available. Most of them only require a function which computes 
the action of the Hamiltonian on any given input vector, as explained below. It 
is important to know that there exist some preconditioners that transform the 
Hamiltonian, making it cheaper to evaluate or increasing the convergence for 
certain diagonalization methods, such as the Jacobi-Davidson. An extensive and 
very pedagogical review about not only hermitian problems, but numerical solving 
of algebraic eigenvalue problems can be found in Ref.~\cite{templev}.

\subsection{Basis states and their ordering}
\label{sec3.1}

In order to identify all the states of the basis, every state needs to have 
an associated label. The basis states should have a known and unique ordering, 
in order to be able to run loops over the vectors of the basis. Computing the 
action of the Hamiltonian on the vectors of the basis has to be as efficient 
as possible. In this work we have used the {\it Ponomarev} ordering~\cite{ponopriv}. 
It provides an efficient way to have all vectors of the basis labelled with a 
single integer ranging from 1 to the exact dimension of the Hilbert space, ${\cal N}_N^M$ .
In the procedure devised by Ponomarev, the mapping between a Fock state and its 
integer label can be carried out in both directions using a few, simple computational 
steps. It builds on a recursive  relation for the dimensions of Hilbert spaces of 
different number of particles, 
\beq
{\cal N}_N^M = \sum_{n=0}^O {\cal N}_{N-n}^{M-1} \quad {\rm with} \ N, M, O>0\,,
\label{eq:rec}
\eeq
where $O$ is the maximum occupancy per site, which sometimes is taken smaller than 
$N$ to speed up the computations. Eq. (\ref{eq:rec}) allows one to devise a counting algorithm 
covering all numbers from 1 to ${\cal N}_N^M$. To perform the mapping, one first needs to 
evaluate all ${\cal N}_n^m$ occurring in Eq. (\ref{eq:rec}).

Once this information has been obtained, the algorithm first re-writes the Fock state, 
determining the occupations of the $M$ orbitals, into an $N$-component array 
$(m_1,m_{2},\dots,m_N)$, where $m_i$ denotes the orbital in which the $i$th atom is. This 
becomes a simple one-to-one map by demanding $m_i \geq m_j$ for $i<j$. The 
integer label of the Fock state, $n_\beta$, is then obtained as
\beq
\label{ponomap}
n_\beta = 1 + \sum_{j=1}^{N} {\cal N}_{j}^{M-m_j}.
\eeq
With this, we can straightforwardly map a Fock state onto an integer label running 
from 1 to ${\cal N}_N^M$. The opposite map is slightly more complicated, as it involves 
an iterative procedure: Given $n_\beta$, we find $m_N$ by determining the largest 
${\cal N}_N^m<n_\beta$. We then identify $m_N=m$, and continue to determine $m_{N-1}$ 
by finding the largest ${\cal N}_{N-1}^m<n_\beta-{\cal N}_N^{m_N}$, and so on.

Let us see some examples. Consider for instance $N=M=O=6$, with the ${\cal N}_n^m$ 
given in Table~\ref{tab:pono}, and the Fock vector $|\beta\rangle=|103020\rangle$. This 
tells us that the first site is occupied by one atom, the third site is occupied 
by three atoms, and the fifth site is occupied by two atoms. Accordingly, we 
re-write this information in agreement to the rule $m_i \geq m_j$ for $i<j$
as $(m_1,m_2,m_3,m_4,m_5,m_6)=(5,5,3,3,3,1)$. Plugging this into Eq.~(\ref{ponomap}), 
the integer label is then found as:
\beq
n_\beta = 1+ {\cal N}_1^1 + 
   {\cal N}_2^1 +  {\cal N}_3^3 + {\cal N}_4^3 + 
  {\cal N}_5^3 + {\cal N}_6^5   = 258.
\eeq
In Table~\ref{tab:pono} we illustrate this mapping graphically for a second example, 
and explain how to operate in the inverse direction, that is from the integer label 
to the Fock state.

\begin{table}
\centering
\begin{tabular}{lr|ccccccc|r}
 &&&&&&&&& $n_{6-M}$ \cr
&6&1&6&21&56&126&252&462& \cr
&5&1&5&15&35&70&\underline{\bf 126}&\underline{\bf 210}&{\bf 2}\cr
&4&1&4&10&20&\underline{\bf 35}&56&84&{\bf 1}\cr
$M$&3&1&3&6&\underline{\bf10}&15&21&28&{\bf1}\cr
&2&1&2&3&4&5&6&7& {\bf 0}\cr
&1&1&\underline{\bf1}&\underline{\bf1}&1&1&1&1&{\bf 2}\cr
&0&1&0&0&0&0&0&0&{\bf 0}\cr
\hline
&0&0&1&2&3&4&5&6& \cr
&&&&&$N$&&&\cr
\end{tabular}
\caption{Number of Fock states for a given $N$ and $M$. The diagram shows the 
procedure to obtain the index for the Fock vector $|\beta\rangle=|211020\rangle$ 
for $N=6$ and $M=6$. The corresponding index is $n_\beta=1+210+126+35+10+1+1=383$ 
out of the 462 states in the Hilbert space. The inverse procedure can also be 
read out, starting with $n_\beta=383$, we look for the largest number in the $N=6$ 
column which is already smaller than $n_\beta$, in this case, $210$, we put one particle 
in the first mode, then we compare the remained with the values in the $N=5$ column, 
turns out larger than 126, and so on. In \ref{appa} we provide explicit 
Fortran codes for the procedures. \label{tab:pono}}
\end{table}
The inverse procedure, to go from the index to the actual Fock state is also 
fairly simple, subroutines coded in Fortran are provided in \ref{appa}.

In our bosonic case, we have used the Fock states of populations of the lattice 
sites, see Eq.~(\ref{eq:FocSt}), allowing up to $N$ particles per site and 
restricting the total number of particles in the system to $N$. For fermions the 
main difference is that the maximum population per site is 1, due to the Pauli 
exclusion principle, the labelling scheme works well simply considering $O=1$ 
in Eq.~(\ref{eq:rec})

\subsection{Use of sparse matrices to store the Hamiltonian matrix}
\label{sec3.2}

Since Hamiltonians are hermitian, roughly half of the entries in the 
matrix are easily derived from the other half. This fact can be used to reduce storage memory, and to 
prevent us from redundant computations. Moreover, Hamiltonians of physical models are 
typically not very dense. In the case of the Bose-Hubbard model, different states in the 
Fock basis are connected through hopping processes, but clearly this leads to non-zero 
matrix elements only between Fock states differing in two entries. 

The most benefits of this sparseness can be made, if the matrix is stored in a sparse 
matrix format. We then only care about the non-zero elements which are stored in three 
1D arrays of length $L$, with $L$ being the number of non-zero elements. Typically, 
${\cal N} < L \ll  {\cal N}^2$, with ${\cal N}$ the Hilbert space dimension. Two of 
these arrays carry the integer labels of the pairs of states which are connected 
by the Hamiltonian (i.e. column and row of every non-zero matrix element). The third 
array stores the complex amplitude of such process, i.e. the value of the corresponding 
matrix element. In the case of the BHM, the length $L$ is bounded from above 
by $\left(1 + M z \right){\cal N}_N^M$, where $z$ is the coordination number. Each Fock state can 
(at most) be connected to $M z$ other states through hopping processes, and to itself 
through the interaction.

\subsection{Geometry of the lattice}
\label{sec3.3}

In our computations we have considered a chain of atoms, but the topology and coordination 
number of the lattice could easily be changed. All information about the lattice is stored 
in an $M\times z$ array of adjacencies $A$. Its elements $a_{\delta}^{i}$ contain, for each 
site $i$, the labels $\delta$ of all neighbouring sites.

This can be extended to any kind of neighbourhood (nearest neighbours, next nearest 
neighbours, superlattices, anisotropic models, fully connected models, \dots). We then 
simply define a generalized array $A$ of dimension $M\times z\times w$. Here, $w$ counts 
the different types of neighbourhoods, and $z$ is the largest coordination number in 
any neighbourhood. For instance, assume a 2D lattice with nearest- and next-nearest-neighbour 
hopping. Each site is then connected to 4 nearest neighbours, as well as 4 next-nearest 
neighbours, thus $z=4$. We have two different types of connections, thus $w=2$. 
Or consider a triangular lattice. In the isotropic case, each site is equally connected 
to six neighbours, e.g. $z=6$ and $w=1$. If the model becomes anisotropic, we have 
three types of connections, $w=3$, to two different sites, $z=2$.

The important advantage of implementing the lattice geometry as described here is 
its flexibility, specially in the implementation on inhomogeneous and anisotropic 
models. The counterpart, it should be said, is that it does not make use of lattice 
symmetries, like translational symmetry in the case of periodic boundaries, or parity 
symmetry for finite lattices. Since the Hamiltonian commutes with the corresponding 
symmetry operator, the Hamiltonian matrix is block-diagonal in the eigenbasis of a 
symmetry operator. The diagonalization can then be performed within each block 
separately. A comprehensive instruction for implementing translational symmetry in 
the exact diagonalization code can be found in Ref.~\cite{0143-0807-31-3-016}. The 
largest block in the translationally invariant eigenbasis has a dimension which is 
approximately by a factor $1/M$ smaller than the full Hilbert space of $N$ bosons on $M$ sites.

\subsection{Diagonalizing the Hamiltonian}
\label{sec3.4}

As mentioned earlier, diagonalization algorithms differ greatly, but all of them need 
to calculate the action of the Hamiltonian onto the basis vectors. In exact methods 
the outcome of this calculation is stored in a matrix, and the unitary transformation 
diagonalizing this matrix is determined numerically. The advantage of the direct method 
is the fact that they provide the full spectrum of the Hamiltonian. However, direct 
methods are only feasible for matrix sizes on the order to $10^4\times 10^4$, e.g. 
7 particles in 10 sites, see Table~\ref{tab:hss}. 

Beyond that, only iterative methods can be employed. Even where direct methods are 
still possible, iterative methods are much faster in providing only a few eigenvalues 
and eigenvectors. Iterative methods repeatedly apply the Hamiltonian on a set of 
state vectors, thereby filtering out an effective subspace. This procedure can be 
designed such that the invariant subspace corresponds to the low-energy subspace. Since 
it is typically much smaller than the total Hilbert space, direct methods can finally 
be used to diagonalize the Hamiltonian within the low-energy subspace. 

While the iterative methods do not require that the action of the Hamiltonian 
on a basis vector is stored in memory, nevertheless this information is 
frequently needed in order to perform the iterative multiplications. Thus, in particular 
if memory restrictions forbid to store this information, it is crucial for these 
algorithms to quickly evaluate the action of the Hamiltonian on a base vector 
``on the fly''. For this goal, the labelling scheme presented above is an important ingredient. 

Let us analyse the different steps the diagonalization algorithm has to go through. 
Consider an arbitrary state represented as a state vector in the Fock basis, 
$\ket{\Phi}=\sum_{\beta}^{{\cal N}_N^M}{c_{\beta} \ket{\beta}}$. The Hamiltonian is applied in two loops:
\begin{itemize}
\item One loop runs through all elements in the Fock basis, 
$i_\beta=1,\dots,{\cal N}_N^M$. In this loop, we perform a map from the state label $i_\beta$ 
onto the occupation numbers.
 
\item A second loop runs through all terms in the Hamiltonian, $\hat{H} =\sum_j  \hat{H}_j$, 
where $\hat{H}_j$ is a  monomial of creation and annihilation operators, e.g. 
$H_j = \hat{a}^\dagger_3 \hat{a}^\dagger_5 \hat{a}_2\hat{a}_{14}$. Clearly, each step in 
this loop maps the state $|\beta\rangle$ onto a new basis state $|\beta'\rangle$, 
with an amplitude $w_j^\beta$:
\begin{equation}
\hat{H}_j \ket{\beta} = w_j^\beta \ket{\beta'}.
\end{equation}
It is straightforward to determine both the new state $|\beta'\rangle$ in the 
occupation number basis, and  the amplitude $w_j^\beta$. Using the mapping from occupation 
numbers onto state labels, we also find $i_{\beta'}$.
\end{itemize}
Accumulating the amplitudes $w_j^\beta$ in the $i_{\beta'}$th component of the new state vector, 
both loops together produce $\ket{\Phi'}=\hat{H}\ket{\Phi}$.
In summary, the main computational task is the mapping between labels and states back 
and forth, and application of monomials to the states. Let us exemplify this for two  
of the monomials in Eq.~(\ref{eq:Ham_phi_zero}). As an initial state we take the 
Fock state 383 from Table~\ref{tab:pono}, e.g. $\ket{\Phi}=\sum_i\Phi(i)\ket{i}=\ket{383}$, 
or $\Phi(i)=\delta_{i,383}$. First, we translate the Fock state into the occupation-number 
basis: $|\beta\rangle=|211020\rangle$. Then, we apply all monomials, e.g.
\begin{eqnarray}
{\frac{U}{2}} \hat{n}_1(\hat{n}_1-1) |211020\rangle &=& {\frac{U}{2}} 2\times 1 |211020\rangle \nonumber\\
-t_{1,2} \hat{a}^\dagger_2 \hat{a}_1 |211020\rangle &=&  = -2t_{1,2} |121020\rangle \,,
\end{eqnarray}
The first monomial corresponds to an interaction term. It is diagonal in the Fock basis 
and thus is easily evaluated. We accumulate on the output vector, 
$\Phi'(383)=\Phi'(383)+ U \Phi(383)$. The second monomial represents a tunnelling 
term, and is not diagonal in the Fock basis, that is, it changes the state. The new 
state and the amplitude can easily be found, and using the Ponomarev mapping, we 
finally identify the label of the new state, $n_{121020}=210+56+35+10+1+1=313$. This means, 
we accumulate the amplitude on that position the resulting vector, 
$\Phi'(313)=\Phi'(313) -2 t_{1,2} \Phi(383)$ in this case.

Once we have this procedure, the iterative methods will perform a number of calls to 
this procedure to obtain approximate values for the desired part of the spectrum. In this 
work we have used the ARPACK package~\cite{citeulike:1296647}, which requires on the order of 
600 calls to this procedure to obtain the first 10 states of the Hamiltonian. With this, 
we are able to obtain the ground state and first excitations of systems 
of up to $5\times10^6$ states. 

\section{Results for the boundary between Mott insulator and superfluid}
\label{sec4}

We are now ready to apply the exact diagonalization method to the Bose-Hubbard model. 
Our goal is to find the value of $t/U$ at which the MI is no longer the GS 
of the system and it starts to be a SF in an infinite system with $N=M$, which is 
known as the critical value of the order parameter of the MI-SF transition at filling $q=1$. Although we also show a few results for the 2D square lattice, our focus is on a homogeneous 1D Bose-Hubbard chain with nearest 
neighbour hopping.
The superfluid to Mott-insulator phase transition exhibited 
by the BHM with a commensurate number of particles, $N/M \in \mathbb{N}$, in 
$d$ dimensions belongs to the $\left(d+1\right)$D $XY$ model universality class. 
For the 1D model, the exhibited phase transition is of the Berezinskii-Kosterlitz-Thouless 
type~\cite{PhysRevB.40.546} (BKT). This phase transition is known to be infinite order ---every derivative of the 
free energy is continuous--- and very sensitive to finite size effects. As we will see in this section, this makes the determination the phase boundary extremely hard.

In order to interpret our numerical results, we will follow three different strategies: In Sec. \ref{sec4.1.1}, we will consider the ground state vectors and determine their overlap with the analytic trial wave functions for the Mott phase and the SF phase. In Sec. \ref{sec4.1.2}, we will analyze the insulating gap which in the thermodynamic limit closes at the transition point. In Sec. \ref{sec4.1.3}, the scaling behaviour of the system is analyzed. We shall stress that all three approaches come with their own limitations, which will be discussed in each subsection. Accordingly, it is also not surprising that each method produces quantitatively different results.

In all calculations, we restrict ourselves to hopping between neighbouring sites $k$ and $j$, $t=t_{k,j}$. 
This keeps the essential symmetries to produce the Mott insulator to superfluid 
phase transition, cf. Ref.~\cite{rey-thesis}. We also take $U=1$, as only the ratio between $t$ and $U$ determines the system behaviour (for $U>0$).

\subsection{Overlap.}
\label{sec4.1.1}

\begin{figure}[t]
\includegraphics[width=\columnwidth]{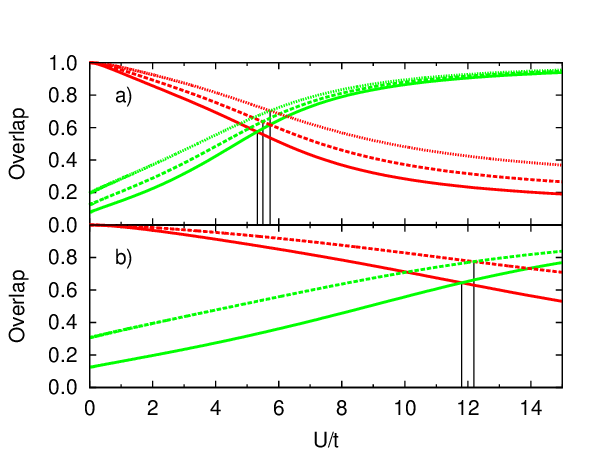}
\caption{a) Overlap of the GS of the system with the analytical SF (red) and MI (green) 
states in 1D lattices with periodic boundary of 5 (dotted line), 6 (dashed line) and 
7 (solid line) sites. b) Computations in 2D: 2x2 (dashed line) and 3x2 (solid line) 
lattices with periodic boundary. The abscissa where the two overlaps have the same value 
is marked to ease visualization. Filling factor $q=1$ so, $N=M$ in all the cases.}
\label{fig:SFiMI_PBC}

\end{figure}
Since we have the eigenstates of the system, which is a quantity that not every method
is able to obtain, we may try to use this information to find the transition value.
Then, we will compare the obtained ground states at different values of $U/t$ with the
analytical solution of the system in the cases $U/t=0$ and $U/t=+\infty$. In
particular, we compute the overlap between GS and trial states as a function of
$U/t$, 
\begin{equation}
OV=|\braket{\Phi_{\rm Analytic}}{\Phi_{\rm GS}}|\,.
\label{eq:OV}
\end{equation}
This overlap is never expected to be zero for finite systems, since
the two trial states become orthogonal only in the thermodynamic limit.
Analytically, we find
\begin{equation}
\left|\braket{\Phi_{\rm MI}\left(q\right)}{\Phi_{\rm SF}} \right|
=\sqrt{\frac{N!}{(M)^N(q!)^{M}}}.
\label{eq:OV_SF_MI}
\end{equation}

Therefore, this method is ill-conditioned for the BKT transition, but we show it for illustrative purposes.
Nevertheless, the overlap $OV$ can estimate the phase boundary by looking for
the value $U/t$ where both overlaps, for the MI and the SF phase, cross each other,
that is, the GS of the system populates them equally, see Fig.~\ref{fig:SFiMI_PBC}.
We denote this value by $(U/t)_{N}$, as it
depends on the number of particles $N$. Performing a finite size
study~\cite{PhysRevA.81.023606}, we estimate the critical value in the
thermodynamic limit, $(U/t)_{\infty}$, by extrapolation.
We assume a size-dependency given by $\left(  \frac{U}{t} \right)_{ \rm M} = A M^{-b} + \left( \frac{U}{t} \right)_{\infty}$, and perform the finite size study for the 1D systems.

This is a naive approach that is routinely used in the study finite-size effects of FQH systems.
The size-dependency is chosen as a power with a variable exponent in place of a linear relation in order to capture any correction depending on non-integer powers.

\begin{figure}[t]
\includegraphics[width=\columnwidth]{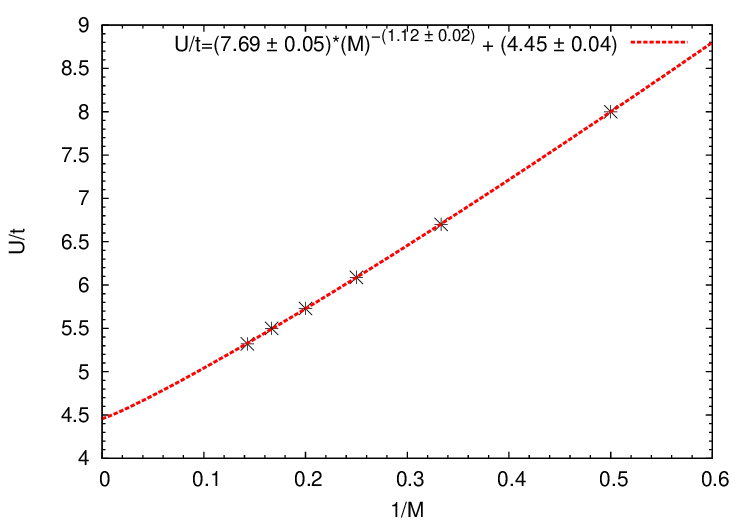}
\caption{Finite-size scaling: The value of $U/t$ at which the crossing of the overlaps happens is plotted as a function of
$1/M$ for a 1D system with periodic boundary. The fitting to the
analytical form, $U/t=a(M)^{-b}+c$  has been made with the non-linear
least-squares Marquardt-Levenberg algorithm. This fit is used to extrapolate to
the thermodynamic limit as explained in the text.}
\label{fig:1D_stud_siz}
\end{figure}

The finite size study is shown in Fig.~\ref{fig:1D_stud_siz}. The extrapolated value for the phase transition in the thermodynamic 
limit is $U/t=4.45\pm0.04$, or, $t/U=0.224\pm0.002$ with a reduced 
$\chi^2=6 \times 10^{-5}$. It is far indeed from most values in the 
literature, cf. Ref.~\cite{PhysRevA.79.013614} for an overview.
The value found here lies between the one from third-order strong-coupling expansion~\cite{PhysRevB.53.2691} and 
the one from density-matrix renormalization-group calculations~\cite{PhysRevLett.65.1765}. 

% This accuracy could be expected, 
% since we have exactly diagonalized the whole Hamiltonian without 
% imposing any restriction on the Hilbert space, thus, without losing 
% any correlations. The only bad feature could be a slow convergence 
% with the size to the value in the thermodynamic limit and the consequent wrong extrapolation.

Thus, based on our knowledge of overlaps in a small system, we are able to predict the phase diagram in the thermodynamic limit, although the overlap itself is certainly not a good figure of merit for the BKT phase transition. In the following subsection, we take the opposite (and more systematic) approach, which characterizes the phase boundary via an order parameter which, in the thermodynamic limit, vanishes exponentially in one of the phases.

\subsection{Insulating gap.}
\label{sec4.1.2}

By means of exact diagonalization, we are able to find the ground state energy of the system with $N$ particles in $M$ sites at a given value of $t/U$,
 $E_{0}\left(t/U,M,N\right)$, in units of $U$, with machine precision.

According to Ref.~\cite{PhysRevB.40.546}, in the phase diagram of the BMH model, the critical value of the MI to SF phase transition is the value of $t/U$ at which the upper and lower boundaries of each Mott lobe cross each other.
We will try to exploit that idea defining an order parameter as the difference in ordinates between the two boundaries as function of $t/U$, following Ref.~\cite{JETPLett.60.177}. 
In the infinite system, that order parameter vanishes for the SF phase, as the boundaries cross each other at the transition value. Meanwhile, it remains finite as long as the GS of the system is the MI state.
At first, we set a definition to find the upper and lower boundaries of the Mott lobes.
According to Ref.~\cite{PhysRevB.40.546}, the upper (lower) boundary of a Mott lobe is given by exciton energy of one particle (hole) in the system.
That is, the chemical potentials of the systems with $M$ sites containing $M+1$ ($M-1$) particles.
Then, we can find the upper (lower) boundary of the Mott lobe at filling $q$ of the system of $M$ sites, $\mu_{M,q}^{+}\left( t/U \right)$ ($\mu_{M,q}^{-}\left( t/U \right)$), as,
\begin{align}
\mu_{M,q}^{+}\left( t/U \right)= & E_{0}\left(t/U,M,q M+1\right)-E_{0}\left(t/U,M,q M\right)\\
\mu_{M,q}^{-}\left( t/U \right)= & E_{0}\left(t/U,M,q M\right)-E_{0}\left(t/U,M,q M-1\right).
\label{eq:Mus}
\end{align}

\begin{figure}[t]
\includegraphics[width=\columnwidth]{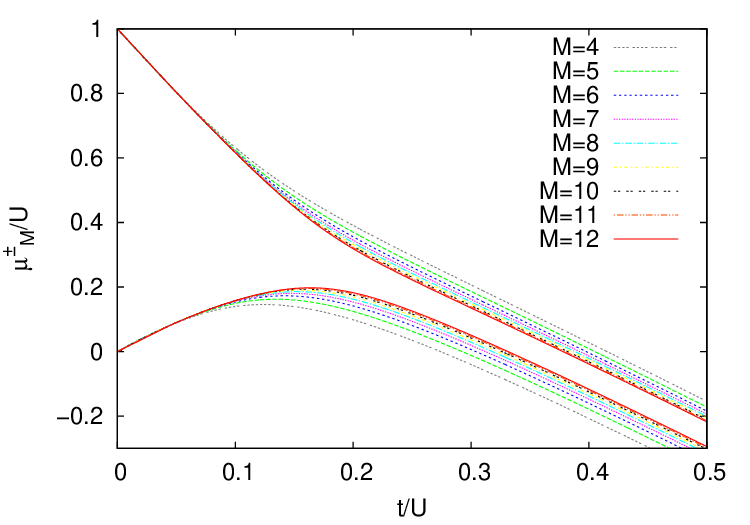}
\caption{Boundaries of the Mott insulator region with $N/M=1$ for finite 
size systems. The sizes are $M=4,5,6,7,8,9,10,11,12$.
The upper family of curves is $\mu_{M}^{+}$ and the lower is $\mu_{M}^{-}$.}
\label{fig:Chem_Pot}
\end{figure}
In Fig.~\ref{fig:Chem_Pot}, the value of $\mu_{M,q=1}^{+}\left( t/U \right)$ and $\mu_{M,q=1}^{-}\left( t/U \right)$
is plotted as a function of $t/U$ for $M=4$ to $M=12$.
This figure shows the famous Mott lobes for finite systems. Notice that for our finite sizes and fixed number of particles, the boundary 
never closes, that is, the upper and the lower boundary of the lobe do not merge. However, it can clearly be seen how these two boundaries approach each other upon increasing the number of particles. 

The energy gap in the MI phase, for any value of $t/U$, corresponds to the particle-hole excitation,
which is the difference between $\mu_{M,q}^{+}\left( t/U \right)$ and $\mu_{M,q}^{-}\left( t/U \right)$ for a fixed $t/U$.
So, we define the single-particle excitation gap of the lobe with filling $q$ in a system with $M$ sites as,
\begin{align}
\begin{split}
\Delta_{M,q}\left( t/U \right)=& \mu_{M,q}^{+}\left( t/U \right)-\mu_{M,q}^{-}\left( t/U \right) \\
=& E_0\left(t/U,M,q M+1\right)+E_0\left(t/U,M,q M-1\right)-2E_0\left(t/U,M,q M\right)\,.
\label{eq:SPEGap}
\end{split}
\end{align}

In the standard quantum phase transitions the single-particle excitation gap is particularly well suited as an 
order parameter because in an infinite system it vanishes in the superfluid phase, meanwhile it remains finite in the MI phase.
Unfortunately, the single particle gap is not well suited to locate the transition in the 1D case. In the BKT transition the gap is exponentially weak near the criticality, hardly detectable in finite systems. Hence, the formula above is by construction incorrect for small gaps in the Mott insulator phase.
In addition, the studied systems exhibit finite size gaps due to the small size. Those gaps may dominate the single-particle excitation gap in the transition and clearly do in the superfluid phase, and besides, they can have different extrapolation exponents than the single-particle excitation gap.
Obviously, a reliable extraction of the gap is also possible from Monte-Carlo methods, and possibly they will do a better job for this transition.
The analysis of the energy gap performed in the present case, leads indeed to the results which do not have a clear physics meaning; nevertheless, one can estimate quite well the position of the criticality from that.

\begin{figure}[t]
\includegraphics[width=\columnwidth]{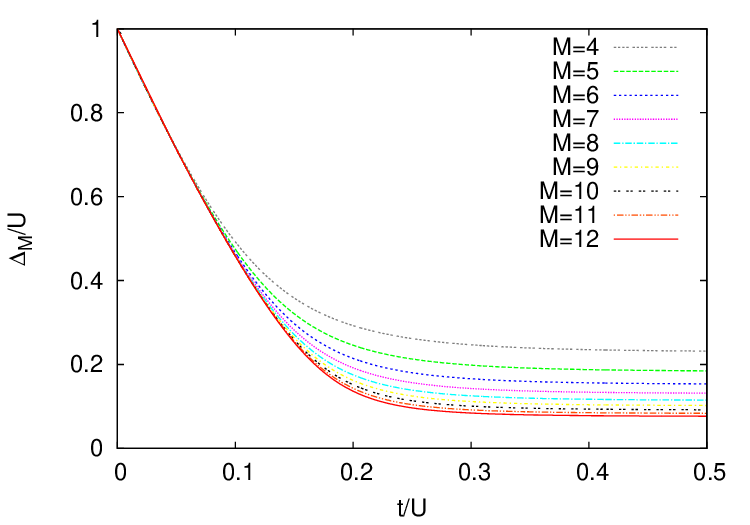}
\caption{Single-particle excitation gap in the regime $q=1$, for finite 
size systems. The sizes are $M=4,5,6,7,8,9,10,11,12$.}
\label{fig:Gaps}
\end{figure}

For simplicity, we define the single-particle excitation gap in the Mott lobe of filling $1$ as $\Delta_{M}\left( t/U \right) \equiv \Delta_{M,q=1}\left( t/U \right)$.
In Fig.~\ref{fig:Gaps}, the value of $\Delta_{M}\left( t/U \right)$ is plotted as a function of $t/U$ for $M$ from $M=4$ to $M=12$. Notice that the gap does not vanish due to the mentioned domination of the finite size gaps in the superfluid phase, at large values of t/U, while the vanishing gap is an intrinsic property of the superfluid in the thermodynamic limit.

In order to determine the value of $t/U$ for which the phase transition takes place, we have used values of $\Delta_{M}\left( t/U \right)$ as the plotted in Fig.~\ref{fig:Gaps} for $M$ from $M=3$ to $M=13$.
We have used here the fitting method from Ref.~\cite{JETPLett.60.177}:
For every value of $t/U$, we fit $\Delta_{M}\left( t/U \right)$ to a fifth-degree polynomial of the inverse of the size, $1/M$. This expression has six fitting parameters.
The constant term of the polynomial is $\Delta_{\infty}\left( t/U \right)$, which corresponds to the single-particle excitation gap of the thermodynamic system ($M \rightarrow \infty$) as function of $t/U$.
Then, the phase transition takes place at the value of $t/U$ for which $\Delta_{\infty}\left( t/U \right)$ just vanishes.
The determination of $\Delta_{\infty}\left( t/U \right)$ through the regression is just a hidden extrapolation to the infinite system.
Following Ref.~\cite{JETPLett.60.177}, the behaviour of the extrapolation to $M \rightarrow \infty$ could imply a non-integer extrapolation exponent that a polynomial expression could not properly capture. In order to extrapolate the proper value of $\Delta_{\infty}\left( t/U \right)$ in the region where the finite size gaps could potentially play a role ($t/U \gtrsim 0.24$), we have used a fitting expression as function of $1/M^{\alpha}$ instead of $1/M$, where $\alpha$ is a positive real exponent. This adds one extra free parameter to the fitting expression.

%Notice that they saturate at some point to a value that depends on the size of the system.

\begin{figure}[t]
\includegraphics[width=\columnwidth]{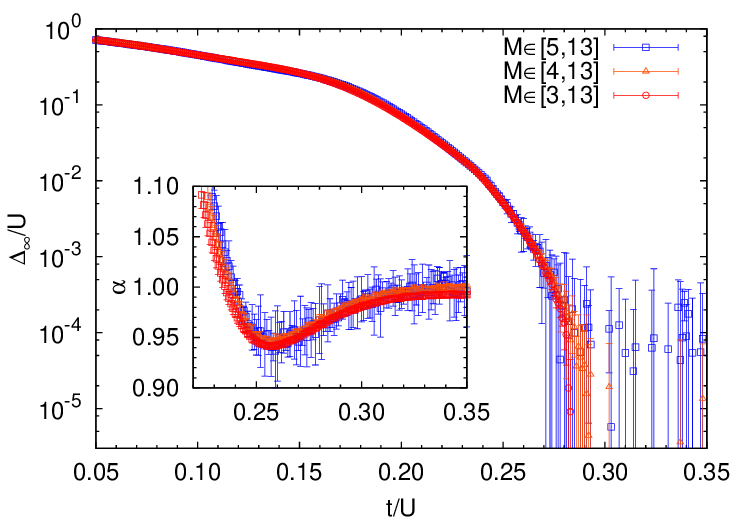}
\caption{Extrapolated value $\Delta_{\infty}$ as a function of $t/U$ 
in a log scale for three different sets of sizes. The inset shows the value of the fitting parameter $\alpha$ as a function of $t/U$ for each set of data.
The errorbars show the $95\%$ confidence intervals.}
\label{fig:LogGap}
\end{figure}

The obtained values of $\Delta_{\infty}\left( t/U \right)$ as function of $t/U$ for three sets of sizes $M \in \{ 3, ..., 13\}$, $M \in \{ 4, ..., 13\}$ and $M \in \{ 5, ..., 13\}$
are shown in Fig.~\ref{fig:LogGap}, along with the corresponding value of the exponent $\alpha$.
The log scale has been used for an easier visualisation of the vanishing point.
In Fig.~\ref{fig:LogGap}, the behaviour of $\Delta_{\infty}\left( t/U \right)$ in units of $U$ is roughly similar for every set of sizes: it starts at $1$ for $t/U=0$ and monotonically decreases to $0$ at $t/U \approx 0.285$. For $t/U \gtrsim 0.285$, the different sets show different behaviours:
The set with sizes $M \in \{3, ..., 13\}$ shows negative, small values of $\Delta_{\infty}\left( t/U \right)$,
while the set with sizes $M \in \{5, ..., 13\}$ shows even smaller, positive and negative values, whose errorbars make them mainly compatible with $0$.
The set with sizes $M \in \{4, ..., 13\}$ shows an intermediate behaviour. It shows positive and negative values of $\Delta_{\infty}\left( t/U \right)$, that are smaller in magnitude than in the former set, but they are more biased to negative values than in the latter set. Some of the values are incompatible with $0$. 
Obviously, any value $\Delta_{\infty}\left( t/U \right)<0$ is clearly unphysical.
Still, the value of $\Delta_{\infty}\left( t/U \right)$ and its dependence on $t/U$ suggest that are a reasonable way to identify the criticality.
The value of $\Delta_{\infty}\left( t/U \right)$ deep in the SF phase is not zero as we know it should, but a negative small value.
This is because we did an extrapolation from small, finite sizes that led to an inaccurate values of the y-intercept, $\Delta_{\infty}\left( t/U \right)$.
As we restrict the analysis to sets of larger sizes, the value of $\Delta_{\infty}\left( t/U \rightarrow \infty \right)$ goes closer to zero, becoming less negative, and even erratic around zero. Consequently, we will treat any small negative value as what it is: an unphysical value that has been obtained just because it is the one that better meets the fitting relation with data from small systems. So, the estimation of the critical value $\left(t/U\right)_c$ will be the value of $t/U$ for which $\Delta_{\infty}$ crosses zero for first time and its uncertainty will be the difference between the latter value and the value of $t/U$ at which the errorbar has crossed zero for first time. 
Then, the obtained critical value for the sets $M \in \{ 3, ..., 13\}$, $M \in \{ 4, ..., 13\}$, and $M \in \{ 5, ..., 13\}$ using this method is $\left(t/U\right)_c=0.285\pm0.002$,$\left(t/U\right)_c=0.292\pm0.006$, and $\left(t/U\right)_c=0.283\pm0.009$ respectively.
Being conservative, we estimate the critical value with this method as the mean of the latter values, weighted with the relative error, giving $\left(t/U\right)_c=0.286\pm0.017$.
Notice that the set of bigger sizes has $8$ different sizes and its data is fitted with an expression with up to $7$ free parameters.
The fact that this system is minimally overdetermined leads to some instability in the values of the fitting parameters and to bigger uncertainties.

The fitting parameter $\alpha$ has remained within the range $\left[0.94,1.00\right]$ for all the values of $t/U$ used in the analysis.
Notice that the transition value of the Ref.~\cite{JETPLett.60.177}, $\left(t/U\right)_c=0.275\pm0.005$, is compatible with ours.
Interestingly enough, our values of the fitting parameter $\alpha$ near the transition are also compatible with their value $\alpha=0.95$.
Also notice the strong discrepancy with the estimation from the previous naiver method.
Despite this method is nothing more than an elaborated extrapolation to infinite size, the final result with this method is within the range of the most recent studies.
It is also compatible with most of values in the literature, due to its broad uncertainty margins.

\subsection{Finite-size effects of the gap.}
\label{sec4.1.3}
We may try to focus in a more general procedure in order to try to get rid of the finite size effects.
The way to proceed in most of phase transitions is the general finite-size scaling hypothesis.
According to it, close to the phase transition, and with the proper finite-size power rescaling of the order and control parameters, 
the curves for different sizes should collapse into a single curve, independent of the size of the system, called universal scaling function. In our case, order and control parameters would be $\Delta_{M,q}$ and $t/U$, respectively.
Regrettably, the exponential closing of the gap characteristic of the BKT transition does not allow such development. 
Since the gap in the superfluid phase closes as $\Delta \sim \exp \left[ -\frac{g}{\sqrt{ | \left(t/U\right)_c - t/U | }} \right]$ ---with $g$ being an unknown constant---, the finite-size corrections become logarithmically small, not potentially as the finite-size scaling hypothesis assumes and therefore, the finite-size power rescaling is not suitable. As a consequence of this behaviour, the BKT transition is known to converge to the thermodynamic limit very slowly when increasing the size of the system.
This is, in order to get rid of finite size effects, order parameter curves corresponding to sizes from a wide range of orders of magnitude are essential.

We have followed an approach similar to the one of the authors of Refs.~\cite{PhysRevA.87.043606} and~\cite{PhysRevB.84.115135}.
They propose an ansatz for the scaling relation of the single-particle excitation gap, 
$\Delta_{M,q}^{\prime} \left( t/U \right) = M \Delta_{M,q} \left( t/U \right) \left[ 1+\frac{1}{2 \ln \left( M \right) +C} \right]$ where $\Delta_{M,q}^{'}\left( t/U \right)$ is the rescaled gap, and $C$ is an unknown constant. Those authors found that $C \rightarrow \infty$ for the standard BHM so, the logarithmic correction becomes negligible.
We defined the rescaled reduced control parameter as $\tilde{t} \equiv \frac{t/U - \left(t/U\right)_c}{\left(t/U\right)_c} M^{a}$, where $a$ is an scaling exponent. The former takes the value $\tilde{t}_c=0$ at criticality.
We also propose the rescaling $\Delta_{M}^{\prime} \equiv \Delta_{M} M^{b}$ for the order parameter, where $b$ is an scaling exponent.
Both, $a$ and $b$ are related to the critical exponents of the universality class of the phase transition. From it we already knew that they should be $a=1/2$ and $b=1$, respectively. Notice that this implies a potential relation that will deviate from the one given by~\cite{PhysRevA.87.043606} for large enough systems.
Although ED does not allow to compute large enough systems to obtain finite-size effect free results, we proceed with the analysis of the obtained results for illustrative purposes.

We use the fact that, at criticality, the order parameter collapses in a single size-independent universal curve
to find the proper exponents and the critical value of the phase transition through a
minimization of the squared differences between curves of different sizes.
Far from the phase transition, the subleading therms overcome the scaling relation and then, the rescaled order parameter depends on the size of the system.
The problem is to determine how far from the phase transition the system starts to exhibit resolvable finite size effects,
and so, which interval of data points has to be taken in consideration for the minimization.
We call $\tilde{t}^-$ ($\tilde{t}^+ $) the lower (upper) limit of that interval.
That is, the curves of the rescaled order parameter follow the same curve in the interval $\left[ \tilde{t}^-, \tilde{t}^+ \right]$ around the criticality.
Then, we define the figure of merit of the minimization as,
\begin{align}
\begin{split}
S \left( \left(t/U\right)_c, a, b \right) = \sum_{M>M^{\prime}} \int_{\tilde{t}^-}^{\tilde{t}^+} 
{\Delta_{M}^{\prime} \left( \tilde{t} \right)- \Delta_{M^{\prime}}^{\prime} \left( \tilde{t} \right) d \tilde{t}}\,.
\label{eq:Sum}
\end{split}
\end{align}
where the integral is calculated numerically over interpolation of the data points with cubic splines.

\begin{figure}[t]
\centering
\includegraphics[width=0.7\columnwidth]{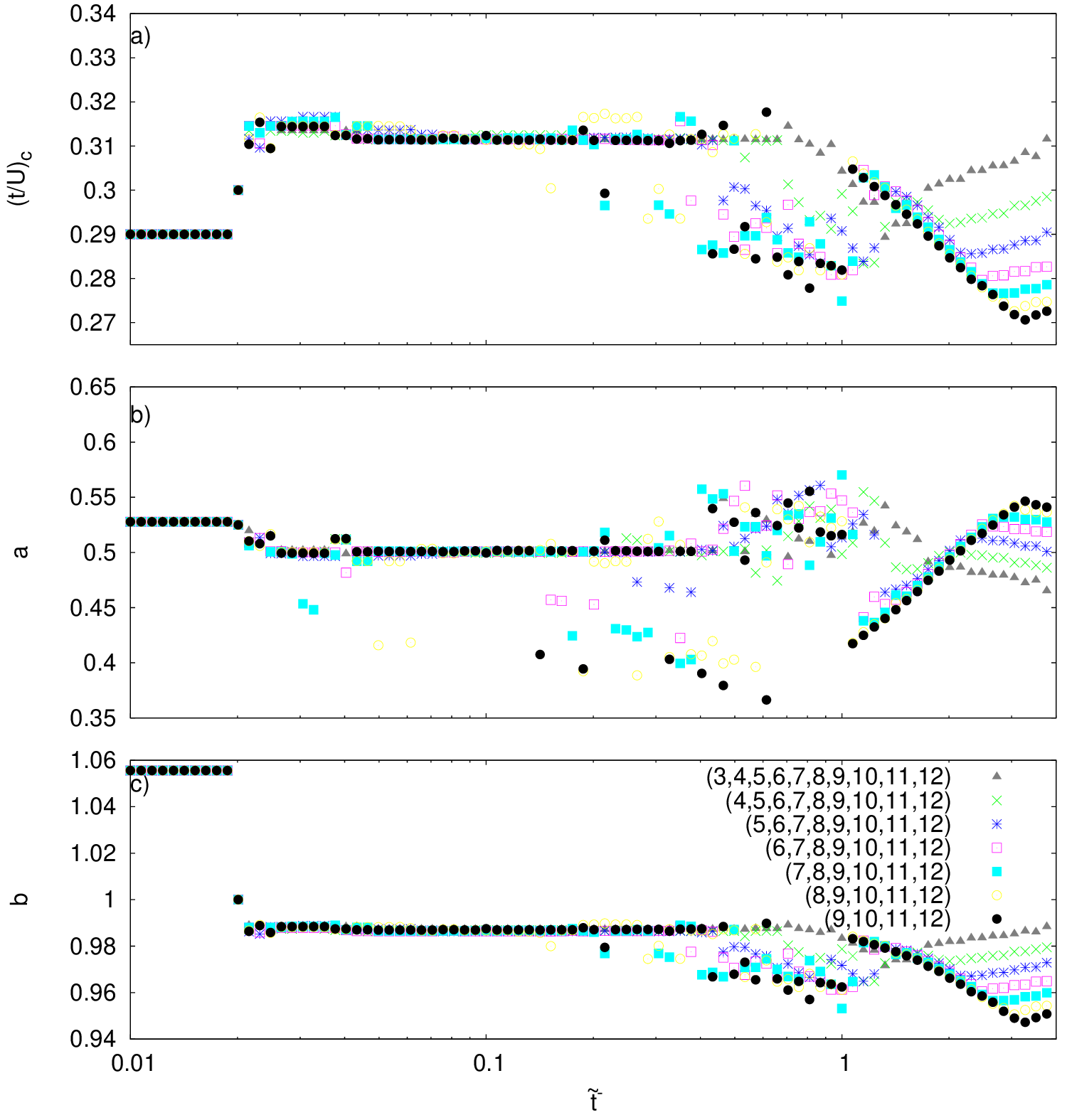}
\caption{Optimal values of $\left(t/U\right)_c$, $a$ and $b$ as a function of $\tilde{t}^-$ to collapse 
several sets of system sizes $M$.}
\label{fig:tcs}
\end{figure}

Since we don't know how far from the critical point the system starts to exhibit resolvable finite size effects,
we try to collapse the curves for several system sizes $M$ as function of $\tilde{t}^-$ and $\tilde{t}^+$ with the following 
procedure:

\begin{itemize}
\item For a given value of $\tilde{t}^-$, we fix $\tilde{t}^+=-\tilde{t}^-/e$, since we have visually 
realized that the lowest values of $S$ are achieved when $\tilde{t}^+ \sim -\tilde{t}^-/2$ holds.

\item We minimize $S$ changing the set of parameters $\left( \left(t/U\right)_c, a, b \right)$. 

\end{itemize}
Then, we find an optimum set of parameters $\left( \left(t/U\right)_c, a, b \right)$ as a 
function of $\tilde{t}^-$. We may expect that when $\tilde{t}^-$ is very small, the number 
of data points is not enough to properly describe the universal scaling function, due to the lack of resolution.
On the other side, when $\tilde{t}^-$ is large enough, the finite size effects play 
a role and the curves are no longer collapsed in the universal scaling function.
This leads to obtaining parameters that are size-dependant and not related to 
the universal scaling function.

For a range of $\tilde{t}^-$ in between, we may expect to have a constant, 
size-independent values of the parameters, showing a plateau. This is due to 
the fact that the curves are collapsed in a universal scaling function, 
which has the same parameters for any choice of $\tilde{t}^-$ and sizes $M$.
In order to control those possible size dependency of the parameters $\left( \left(t/U\right)_c, a, b \right)$, 
we have computed those parameters taking in account different sets of curves: 
pairs of consecutive sizes ($M=11$ and $12$, $9$ and $10$, $7$ and $8$, ...), subsets of the 
larger systems (from $M=9$ to $12$, from $8$ to $12$, ...) and for all of them.

%\begin{figure}[t]
%\caption{Optimal values of $a$ as function of $t^{-}$ to collapse several sets of system sizes.}
%\label{fig:as}
%\end{figure}

%\begin{figure}[t]
%\caption{Optimal values of $b$ as function of $t^{-}$ to collapse several sets of system sizes.}
%\label{fig:bs}
%\end{figure}

\begin{figure}[t]
\includegraphics[width=\columnwidth]{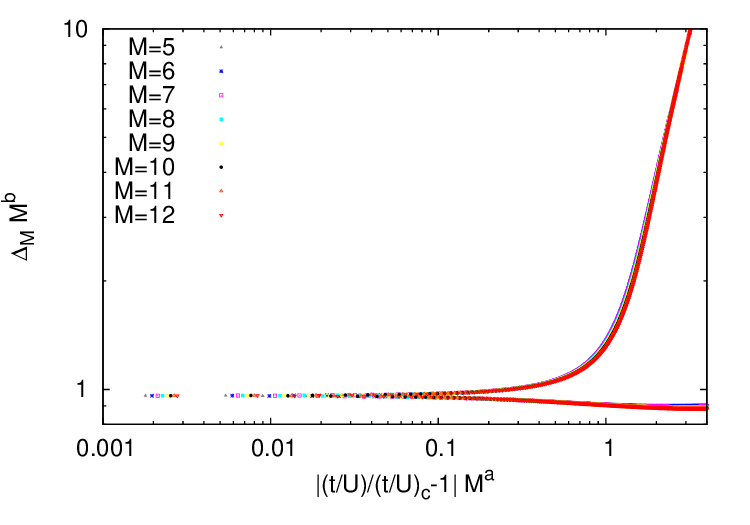}
\caption{Collapse of the curves for $M=5$,$6$,$7$,$8$,$9$,$10$,$11$, and 12 sites 
for the estimated parameters $\left(t/U\right)_c=0.3115$, $a=0.501$, and $b=0.987$ in a log-log scale.}
\label{fig:collapse}
\end{figure}

The parameters $ \left(t/U\right)_c$, $a$, and $b$ for a several size sets are shown in 
Fig.~\ref{fig:tcs}. According to those results, the estimated values are: 
$\left(t/U\right)_c=0.3115\pm0.0010$, $a=0.5010\pm0.0010$, and $b=0.9870\pm0.0010$. 
The fact that the parameters that we have found do not have a resolvable 
size dependency seems quite noticeable. It is because our set of sizes 
are too clustered to resolve the differences due to the size. Notice that we 
have let both exponents, $a$ and $b$, to vary, despite we know their value. This allows 
to explore a broader area of the space of parameters to improve the final value 
of $\left(t/U\right)_c$, and let the minimization find the proper scaling exponents by itself.
Additionally, it gives us a proof of the goodness of the scaling. As a matter of fact, 
the value of the exponent $b$ is several error bars 
below the expected value $b=1$. It is due to the fact that the small sizes we studied didn't allowed to get rid of the finite-size effects. Then, the analysis has led to a non-universal coefficient. Reminding that the size corrections in the BKT transition are logarithmic becomes clearer that the set of sizes shall include sizes with larger orders of magnitude.
It has to be stated that potential scaling relations are wrong for analysing the BKT transition, 
but with this treatment a good value is fortuitously obtained because of the small sizes studied ---given that the value obtained for the exponent $b$ does not correspond to the expected, $1$.
Finally, the collapse of various system sizes with those parameters is shown in Fig.~\ref{fig:collapse}.

\subsection{Summary}
Given that the most recent numerical results localize the BKT transition at values $t/U$  between $0.26$ and $0.31$, we must clearly state that our first approach considering the overlaps fails, as it yields $(t/U)_{\rm crit}=0.224\pm0.002$.  Despite the nature of the BKT and the weakness of the gap even in the insulating phase, the second method produces a result which agrees with the literature, $(t/U)_{\rm crit}=0.286\pm0.017$. Also our third approach, the scaling analysis, produces a result which is still compatible with the literature, $0.3115\pm0.0010$, although the underlying scaling hypothesis does not hold for the BKT transition.

\section{Beyond the standard BHM}
\label{sec5}

A number of modifications to the standard Bose-Hubbard model have been studied.
Those modifications include different topologies and coordination numbers of the lattice,
inhomogeneous potentials, negative interactions, additional neighbouring interactions,
long range interactions, among others.
Exact diagonalization very suitable for most of those modifications,
due to the lack of assumptions on the parameters.
We have played with a couple of modifications:
inhomogeneous lattices, and attractive on-site interactions.

\subsection{Phase transitions in a deeply biased lattice}
\label{sec5.1}

An interesting modification of the SF to MI transition is obtained by 
considering a lattice with a large attractive bias. In this case the tendency to form a
superfluid is suppressed, as in the limit of weak interactions the particles prefer
to localize on the biased site. Increasing repulsive interactions, the system 
reaches the Mott phase, undergoing several transitions in which the number 
of particles on the biased site is reduced by one. The large inhomogeneity 
is produced by making the potential energy in the $k$th site much lower 
than the others. Theoretically, we take it into account by adding the 
term $-\epsilon \sum_i^{M}{\hat{n}_{i}\delta_{i,k}}$ to the Bose-Hubbard 
Hamiltonian.

To evaluate the effect of the bias potential in the system, we 
introduce the fluctuation of the number operator in the $i$th place,
\begin{equation}
(\Delta \hat{n}_i)^2=\expval{(\hat{a}_i^{\dagger}\hat{a}_i)^2}-
\expval{\hat{a}_i^{\dagger}\hat{a}_i}^2.
\label{eq:Nfluct_Def}
\end{equation}
It can be written explicitly with the number operators in the Fock 
basis. Moreover, due to the fact that the Fock states are 
eigenstates of $\hat{n}_i$, the only nonzero contribution occurs 
when $\ket{\beta^{\prime}}=\ket{\beta}$. So,
\begin{equation}
(\Delta \hat{n}_i)^2=\sum_{\beta}|c_{\beta}|^2\expval{\hat{n}_i}_{\beta}^{2}
-\left[\sum_{\beta}|c_{\beta}|^2\expval{\hat{n}_i}_{\beta}\right]^2,
\label{eq:Nfluct}
\end{equation}
where $\expval{\hat{n}_i}_{\beta}$ means $\bra{\beta} \hat{n}_i \ket{\beta}$. 
The fluctuation of the on-site number of particles may serve as a precursor of a
phase transition which involves redistribution of the particles in
the ground states. In the presence of a strong bias potential, $\epsilon \gg
t$, several peaks of the number fluctuations occur upon tuning $U/t$.

\begin{figure}[t]
\includegraphics[width=\columnwidth]{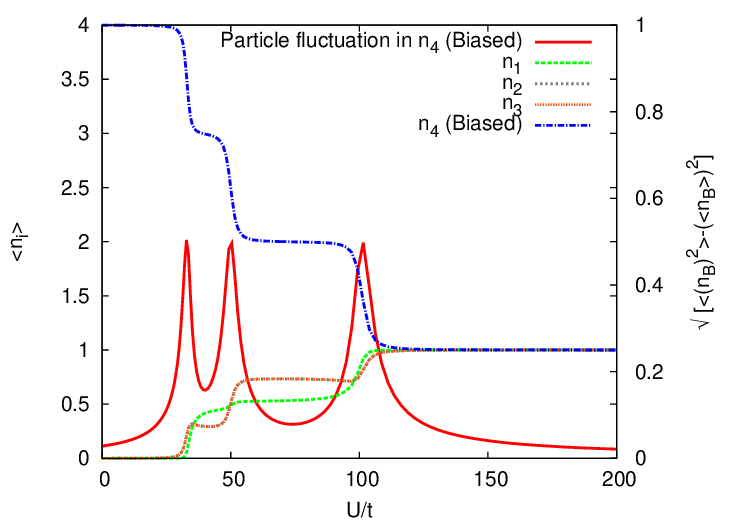}
\caption{Transition from the fully localized state to the MI phase 
in a deep biased 2x2 lattice with open boundary condition. The bias is taken to
be 
$\epsilon = 100t$ in the $4$th site. The values of the average population 
of all sites is depicted together with the fluctuation of the number 
of particles in the biased site (red solid curve). The direct hopping 
between the $4$th site and the $1$st is not allowed and hopping between 
the $4$th and the $2$nd and $3$rd are equivalent. Note the 
clear peaks in the number fluctuation for fixed values of $U/t$ corresponding 
to the transitions described in the text.}
\label{fig:fluctioccnums}
\end{figure}

In Fig.~\ref{fig:fluctioccnums}, we chose $\epsilon=100t$, and study a square
lattice consisting of a single plaquette, that is, four sites. Accordingly, 
we observe $N-1=3$ peaks of the number fluctuations upon tuning from $U/t=0$ 
to large values of $U/t$. In order to infer which mechanisms produces the 
fluctuations, we have calculated the population of each site in the lattice, 
simply by taking  the diagonal values of the OBDM, plotted in
Fig.~\ref{fig:fluctioccnums}. When the fluctuation reaches a maximum, the
population in the biased site decreases by one. Between two consecutive 
fluctuation peaks, the populations remain mainly constant, showing 
plateaus with a step structure. The last peak of the fluctuations, occurring at
the largest value of $U/t$, indicates a transition into the MI phase: We find
that for larger values of $U/t$, the population of all the sites takes the same
integer value $q$, and the fluctuation decrease monotonically to zero.

The values of $U/t$ for which fluctuation maxima appear can be parametrized by $U/t=100/i$, for
$i=1,\cdots,N-1$. These values are easily explainable for the MI with $q=1$,
keeping in mind the Hamiltonian in Eq.~(\ref{eq:Ham_phi_zero}): the migration 
happens when the energy of keeping the particles in the same site becomes 
greater than extracting one particle from the biased site to 
place it in other site without particles,
\begin{equation}
\frac{U}{2} n_{B}(n_{B}-1) -\epsilon n_{B}=\frac{U}{2} (n_{B}-1)(n_{B}-2) -\epsilon (n_{B}-1)\,,
\label{eq:conta}
\end{equation}
where we have neglected the hopping term $t$, which is small compared to $\epsilon$ 
and $U$. The subindex $B$ denotes the biased site. From this equation, we obtain 
the condition,
\begin{equation}
U=\frac{\epsilon}{n_B-1}\,,
\label{eq:troba}
\end{equation}
where $n_B$ is a positive integer which $1<n_B \leq N$.

As can be seen in Fig.~\ref{fig:fluctioccnums}, in general the unbiased sites
are not equally populated. When the interaction is large enough to expel the
first particle from the biased site, the second most populated site is the one 
which is not directly connected to the biased site. This might appear 
counterintuitive in the first place, but one has to bear in mind that a 
particle on this site benefits from having two empty neighbours, allowing to 
reduce energy by tunnelling processes to these sites. On the other hand, once 
a second particle is pushed out from the biased site, the situation changes, 
and two nearest neighbours of the biased site become more populated. But now, 
two particles occupying these two sites still can share the empty neighbouring 
site for virtual tunnelling.

\begin{figure}[t]
\includegraphics[width=\columnwidth]{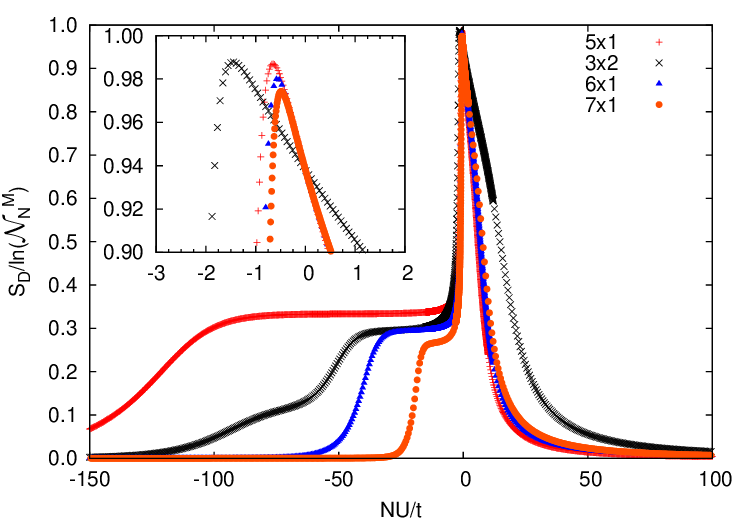}
\caption{Entropy $S_D$ of the GS in a system with attractive interactions,
for 5 to 7 particles in different geometries with periodic boundaries. The plot is
zoomed in order to  appreciate the weakly attractive regime. In particular, it
is worth emphasizing the  fact that the maximum of the entropy, maximal
delocalization in Fock space, is not 
achieved for zero interaction but for slightly attractive one. The bias is $\epsilon=10^{-10}t$.}
\label{fig:VN_PBC}
\end{figure}

\subsection{Attractive interactions: Localization}
\label{sec5.2}

As studied for the two-site case in Refs.~\cite{PhysRevA.57.1208,PhysRevA.81.023615},
systems with attractive interactions feature large quantum superpositions due to the 
several competing single-particle ground states~\cite{PhysRevA.74.033612}. 

For $U/t=-\infty$, all the particles in the system will aggregate in a single 
site, so the GS is the Fock state with $N$ particles in the $i$th site 
and $0$ in the other sites. But this state is $M$-degenerate. 
Due to this degeneracy, the ground state can be a superposition of these 
$M$ states. Each one of them aggregates the system in one different 
site of the lattice. In this state, when a particle is fixed in one site, 
all the rest cluster there. So, this state is highly correlated. For the two 
site case, the ground state build a so-called NOON state~\cite{PhysRevA.57.1208}.

In any practical implementation there will be small imperfections that will 
trigger small biases between the sites. It is thus expected, that for sufficiently 
large attractive interactions in realistic systems, the GS will be unique with 
all particles clustered in one site. To account for such effects, we consider 
a slightly biased case which favours one site, the $k$th. 

The localized condensate (LC) state in the $k$th site of the lattice, reads,
\begin{equation}
\ket{\Psi_{\rm LC}(k)}={\frac{1}{\sqrt{N!}}}(\hat{a}_k^{\dagger})^N\ket{0}.
\label{eq:LC}
\end{equation}
In this state, as in the MI, the number of particles in each site is well 
defined and the correlation length vanishes. Different from the MI, also the energy gap vanishes, 
and its value is given by the value of the bias. Since this state is a 
single state of the Fock basis with all the particles localized in the 
same site, the values of $S_1$ and $S_D$ are both $0$.

It is noticed that if several sites on the lattice were biased significantly 
more than the rest, it could be possible to obtain a fragmented condensate. 
It is also possible to engineer the number of fragmented fractions by setting 
a number of biased sites in the lattice.

To understand the system behaviour for intermediate values of the attractive interactions, 
we apply exact diagonalization and calculate the entropy $S_D$ as function of $NU/t$. The 
results are depicted in Fig.~\ref{fig:VN_PBC}. The entropy has its maximum in the 
attractive regime, not at $U/t=0$ where the entropy $S_1$ exhibits a minimum. This 
observation implies that the GS of a weakly attractive system is more uniformly 
distributed over the Fock basis than the GS of the SF phase. Increasing the attractive 
interaction, but keeping the bias smaller than the gap, the system is in a cat-like 
state, with $S_D = \ln\left( M\right)$.
By cat-like state we mean a superposition state of events that mutually exclude each other from happening simultaneously, 
in this case, the superposition of clustering all the particles in every site of lattice.
Finally, for even stronger attraction, the gap 
becomes smaller than the bias. Then the bias term dominates and the system 
localizes on a single site, with a single Fock state being the ground state. 

The phenomenon is similar to the one studied in Ref.~\cite{PhysRevA.81.023615}. 
There, the system is found to go from a binomial distribution in Fock space, 
to a very homogeneous one at slightly attractive interactions. Further increasing 
the interactions, the distribution does not become more homogeneous, but instead 
starts to develop peaks around each of the two-sites, which corresponds to the 
two superposed states of the cat-like structure. In presence of a small bias, 
further increasing the attractive interaction, the system localizes.

%That can be shown in Fig.~\ref{fig:VN_PBC}, where for increasing sizes the localization in a 
%single site happens at a weaker attraction.

Effects in the weakly attractive regime in higher dimensions than 1D are finite-size effects, 
since in the thermodynamic limit, a soft-core system of bosons collapses at any finite value of attractive interactions \cite{Abdullaev2008}.
In 1D, due to the interplay between the kinetic energy and the attractive interaction energy, 
bright soliton solutions arise from the Gross–Pitaevskii equation \cite{Gordon:83}.

Notice that in the weakly attractive regime, the number of populated Fock states 
increases when interactions are strengthened, but the distribution becomes less 
uniform. This behaviour is more pronounced in the cases with open rather than periodic
boundary conditions, as open boundary provide a natural bias with less connected 
sites at the edge of the system.

\subsection{Exact Diagonalization for other problems: quantum Hall physics}
\label{sec6}

When it comes to studying Bose-Hubbard models with Exact Diagonalization, the reader has to notice that,
despite its insurmountable size limitations, one strength of the method is its applicability to a wide range of problems.
As example, just adding complex values to the tunneling, models with gauge potentials can be studied.

In this section, we will briefly outline how the method can also be applied to continuum systems. As an example, we choose the fractional quantum Hall effect, which can be exhibited by fermionic particles (electrons), but also by bosons, e.g. a cold gas of bosonic atoms rotating around the ${\rm z}$ axis in 2D \cite{doi:10.1080/00018730802564122}. In this bosonic scenario, we shall find some analogies to the treatment of the Bose-Hubbard model.

The first step for treating the problem by exact diagonalization again is to construct a basis for the Hilbert space. 
In the quantum Hall effect, the single-particle energy levels are the Landau Levels (LLs), and it is usually enough to consider only one LL, for bosons the lowest LL (LLL).
All states in the LLL are degenerate, and can be labelled by a quantum number $l\geq 0$, the angular momentum along the rotation axis. These angular momentum eigenstates play a role analogous to the sites in the Bose-Hubbard model, and it allows to map between the basis for the Bose-Hubbard model onto the basis of bosons in the LLL. Since, in principle, there are infinitely many single-particle states, though, we have to truncate the basis at a sufficiently large $l=l_{\rm max}$. Due to rotational symmetry, the total angular momentum $L$ along $z$ is conserved. This provides a natural value $l_{\rm max}=L$ for truncating the Hilbert space, but in practice the available angular momentum will be distributed more equally between all particles, so $l_{\rm max}$ can be chosen much smaller, at the order $l_{\rm max} \sim L/N$ for $N$ bosons.

In contrast to the Bose-Hubbard model, due to the degeneracy of single particle levels in the fractional quantum Hall problem, there is no single-particle term in the Hamiltonian. 
Taking into account a trapping potential only introduces a $L$-dependent energy shift. The interactions, though, are much more difficult to treat than in the Bose-Hubbard model, as two particles at $l$ and $l'$ may scatter to arbitrary orbitals $(l+l')/2+x$ and $(l+l')/2-x$. The interactions may lift the huge single-particle degeneracy, and may give rise to a unique state describing a fractional quantum Hall phase.
In order to interpret the numerical results, one tries to identify the fractional quantum Hall phases by scanning through different values of $L$, searching for pronounced gaps. Similar to our strategy presented in Sec.\ref{sec4.1.1}, one can then compare the numerical ground state with trial wave functions by evaluating their overlaps.

In practical applications, the number of particles is clearly restricted to a small numbers, $N \lesssim 20$.
The studies of mixtures of multicomponent systems restricts the computations to even smaller numbers.
For those systems, a subspace containing every Fock-Darwin state of every species \cite{PhysRevB.89.045114} is constructed.
The total Hilbert space is direct sum of the subspaces, 
and hence, the total dimension of the space is the product of dimensions of those subspaces.

\section{Conclusions}
\label{sec7}

We have provided a comprehensive study of Bose-Hubbard models composed of 
a small number of atoms, $\simeq 10$ populating a small number of 
sites, $\simeq 10$. First, we have introduced the Bose-Hubbard model together 
with a detailed description of the exact diagonalization technique employed. 
Then we have concentrated in the Mott insulator to superfluid transition, first 
discussing its characterisation by means of exact overlaps with trial wave functions 
and secondly by performing finite size scaling of the gap. 

We have also studied a highly biased lattice, in which one site is 
considerably deeper than the others. In this case, the system undergoes 
several transitions, from a fully localized state to a MI phase, going through 
partial superfluid phases, in which more and more atoms delocalized prior
to localizing in the MI. The way the MI phase grows in population 
has been shown to proceed stepwise as the interaction is increased. 

In the attractive interactions case, we have considered a small biased 
case, to understand the competition between attraction and 
localization. For sufficiently large attractive interactions, the system 
fully localizes due to the bias. At lower attractions, the 
system develops a cat like structure. Prior to this, the system 
goes through a state in which the number of populated Fock states 
is maximal. 

\ack
BJ-D thanks A V Ponomarev for sharing his labelling routines. This work has been funded by a scholarship from the Programa M\`asters d’Exce\lgem\`encia of the Fundaci\'o Catalunya-La Pedrera, EU grants (EQuaM (FP7-ICT-2013-C No. 323714), OSYRIS (ERC-2013-AdG No. 339106), SIQS (FP7-ICT-2011-9 No.  600645), and QUIC (H2020-FETPROACT-2014 No.  641122)), Spanish Ministerio de Econom\'ia y Competitividad grants (Severo Ochoa (SEV-2015-0522), FOQUS (FIS2013-46768-P), and FISICATEAMO (FIS2016-79508-P)), Generalitat de Catalunya (2014 SGR 401, 2014 SGR 874, and CERCA program), and Fundaci\'o Cellex. B J-D is funded by the Ram\'on y Cajal program.

\appendix
\section{Subroutines for the labelling procedure}
\label{appa}

Explicit Fortran subroutines to generate the Fock basis labelling 
as explained in Sect.~\ref{sec3.1}. First we need to build the Pascal triangle, depending
on the total number of sites and particles, this is done with {\bf buildpascal}. Once 
this is generated, we can use {\bf b2in} and {\bf in2b}, to from the basis to the index or 
vice versa, respectively. 

\begin{verbatim}
c original from A. V. Ponomarev (2009)
      subroutine buildpascal
c lc=number of sites +1
c nc=number of atoms +1
        parameter (lc=4,nc=3)
 
       double precision jbc
       integer cnkc(lc,nc)
       integer jmax
       common/pascal/jmax,cnkc

c builds the rotated pascal triangle
      do i = 1,lc
         cnkc(i,1) = 1
      end do
      do i = 1,lc
         do j = 2,nc
            cnkc(i,j) = 0
         end do
      end do
      do in1 = 2,lc
         cnkc(in1,2) = sum(cnkc(in1-1,1:2))
         if (nc-1.gt.1) then
         do in2 = 1,nc
          cnkc(in1,in2) = sum(cnkc(in1-1,1:in2))
         end do
        end if
      end do 
      jmax = cnkc(lc,nc)
       end
\end{verbatim}

\begin{verbatim}
c ---------------------------------------------
c Returns the many body state bi at position in
c ---------------------------------------------
c original from A. V. Ponomarev (2009)
      subroutine b2in(bi,in)
      implicit none
      integer in,lc,nc,jmax,ind_L,ind_N,indi,k,is,i
      parameter (lc=4,nc=3)
      integer cnkc(lc,nc),bi(lc),suma,M,in1,in2
      common/pascal/jmax,cnkc
c builds the rotated pascal triangle


      in=1
      do indi=1,lc-2
        do ind_N=0,bi(indi)

        if (bi(indi)-ind_N.gt.0) then 
           suma=0.
            do k=1,indi-1 
            suma=suma+bi(k)
            enddo
        if (lc-indi.gt.0.and.nc-ind_N-suma.gt.0) then 
            is=0
           in=in+cnkc(lc-indi,nc-ind_N-suma)
        endif
        endif
        enddo
      enddo

      end     
\end{verbatim}

\begin{verbatim}
c ---------------------------------------------
c Returns the many body state bi at position in
c ---------------------------------------------
c original from A. V. Ponomarev (2009)
      subroutine in2b(in,bi)
      implicit none
      integer in,lc,nc,jmax,ind_L,ind_N,indi
      parameter (lc=4,nc=3)
  
      integer cnkc(lc,nc),bi(lc)
      common/pascal/jmax,cnkc

         indi = in-1
         bi = 0
         ind_L = lc-1
         ind_N = nc
         do while(ind_N.ne.1)
            if(indi.ge.cnkc(ind_L,ind_N)) then
                indi=indi-cnkc(ind_L,ind_N)
                bi(lc-ind_L)=bi(lc-ind_L)+1
                ind_N = ind_N-1                
            else
                ind_L = ind_L-1
            end if
         end do
      end     
\end{verbatim}
\section*{References}
%\bibliographystyle{iopart-num}
%\bibliography{bib}

\providecommand{\newblock}{}

\end{document}